\documentclass[5p,fleqn,sort&compress]{elsarticle}
\usepackage[english]{babel}
\usepackage{hyphenat}
\usepackage[utf8]{inputenc}	
\usepackage{savesym}
\usepackage{amsmath}
\savesymbol{iint}
\usepackage{txfonts}
\restoresymbol{TXF}{iint}
\usepackage[T1]{fontenc}	
\usepackage{graphicx}
\usepackage{siunitx}	
\usepackage{url}		 
\addto\captionsenglish{}

\usepackage{hyperref}
\hypersetup
{   pdfsubject={Nuclear Physics},
	pdfauthor={Jonas Refsgaard}
    pdftitle={Interpretation of the Hoyle state decay},
    pdfstartview=FitH,
    colorlinks=true}
    
\makeatletter
\begingroup
\catcode`\_=\active
\protected\gdef_{\@ifnextchar|\subtextup\sb}
\endgroup
\def\subtextup|#1|{\sb{\textup{#1}}}
\AtBeginDocument{\catcode`\_=12 \mathcode`\_=32768 }
\makeatother

\usepackage{tikz}
\usetikzlibrary{arrows}
\usetikzlibrary{positioning}

\usepackage{siunitx}
\usepackage{physics}

\usepackage{cancel}

\usepackage{pdfpages}

\usepackage[]{csquotes} 
\usepackage{bm}

\usepackage{amssymb}
\usepackage{wrapfig,lipsum,booktabs}

\begin{document}

\begin{frontmatter}

\title{Exclusive decay study of the 16.62\,MeV (2$^-$, T=1) resonance in $^{12}$C }

\author[au]{M. Kuhlwein}
\author[au]{K. Lytje}
\author[au]{H.\,O.\,U.~Fynbo \corref{cor}}
\ead{fynbo@phys.au.dk}
\author[au]{A. Gad}
\author[au]{E. Jensen}
\author[dalh]{O.\,S.~Kirsebom}
\author[au]{M. Munch}
\author[stmary]{J.~Refsgaard}
\author[au]{K. Riisager}

\address[au]{Department of Physics and Astronomy, Aarhus University, DK-8000 Aarhus, Denmark}
\address[stmary]{Department of Astronomy and Physics, Saint Mary’s University, Halifax, Nova Scotia, B3H 3C3 Canada, and TRIUMF, 4004 Wesbrook Mall, Vancouver BC, V6T 2A3 Canada}
\address[dalh]{Institute for Big Data Analytics, Dalhousie University, Halifax, NS B3H 4R2, Canada} 

\begin{abstract}
  The 3$\alpha$ decay of the 16.62\,MeV (2$^-$, T=1) resonance in
  $^{12}$C has been studied for nearly a century starting with one of
  the first nuclear reaction studies at the Cavendish Laboratory in
  the 1930s. In the hitherto latest study published a decade ago a
  model based on earlier work from the 1960s was found to give a good
  account of a set of inclusive data. This model describes the decay
  as an l=3 $\alpha$-particle populating the 2$^+$ state of
  $^8$Be. Here we provide new exclusive data on the 3$\alpha$ decay of
  the 16.62\,MeV resonance, and demonstrate that the decay is best
  described by a model with predominantly l=1 emission with an
  admixture of l=3.
\end{abstract}


\end{frontmatter}


\section{Introduction}
The 16.62\,MeV (2$^-$, T=1) resonance in $^{12}$C is positioned 663\,keV
above the proton threshold in $^{12}$C. With a width of 300\,keV the
state is prolifically produced by impinging $<$1\, MeV protons on
$^{11}$B, which was one of the first nuclear reactions to be studied
at the Cavendish
Laboratory~\cite{Cockcroft:1932vn,Oliphant:1933uq,Dee:1936fk}.

There are at least three different interests in the p+$^{11}$B
reaction at the energies relevant for the population of the 2$^-$
resonance. Firstly, roughly 50\% of the decay leads to the
3$\alpha$-final state (with the rest being back to p+$^{11}$B ), and
the mechanism of this three-body decay has been in focus since the
early work at the Cavendish Laboratory. Secondly, the
$^{11}$B(p,3$\alpha$) reaction is considered a candidate for fusion
energy generators~\cite{Moreau:1977aa}, a possibility which is still
being pursued in research~\cite{Giuffrida:2020aa} and by several
commercial companies. Finally, the reaction is considered as a
potential way to enhance proton cancer therapy efficiency by adding a
concentration of boron to the tumor volume, which will then produce
high-LET $\alpha$ particles when protons in the Bragg peak enter the
tumor and interact with the boron, e.g.~\cite{Cirrone:2018aa}.

Here we present a new complete kinematics measurement of the 3$\alpha$-decay 
of the 16.62\,MeV resonance and demonstrate how conflicting conclusions on its 
decay mechanism from previous studies throughout almost a century are a 
consequence of previous analyses being based primarily on inclusive
data.

\section{Review of previous work}
\label{review}
\begin{figure}[htpb]
	\centering
	\makebox[=0.5\columnwidth][c]{
\begin{tikzpicture}[
		scale=0.35,
		level/.style={thick},
		trans/.style={thin,->,shorten >=2pt,shorten <=2pt,>=stealth,blue},
	]
	\draw[level] (-21cm,15.96cm) -- (-16cm,15.96cm) node[midway,below] {$p+{}^{11}$B} node[above left] {$15.96$};
	\draw[trans] (-16cm,15.96cm) -- (-14cm,16.57cm) node[midway,above=0.5cm] {\scriptsize$ E_{p} = \SI{675}{\kilo\electronvolt} $};

	\draw[level]
		(-14cm,16.57cm) node[above right] {$2^-$} -- (-9cm,16.57cm) node[midway,below] {${}^{12}$C} node[above left] {$16.57$};
	\draw[trans] (-9cm,16.57cm) -- (-7cm,10.39cm) node[midway,right] {$\alpha_{1} $};
	\draw[trans] (-9cm,16.57cm) -- (-7cm,7.36cm) node[midway,left] {$\cancel{\alpha_{0}} $};

	\fill[black!30!white] (-7cm,8.89cm) rectangle (-2cm,11.89cm);
	\draw[level] (-7cm,10.39cm) node[above right] {$2^+$} -- (-2cm,10.39cm) node[above left] {$10.39$};
	\draw[level] (-7cm,7.36cm) node[above right] {$0^+$} -- (-2cm,7.36cm) node[midway,below] {$ \alpha + {}^{8}$Be} node[above left] {$7.36$};
	\draw[trans] (-2cm,10.39cm) -- (-0cm,7.27cm) node[midway,right] {$\alpha + \alpha $};
	\draw[trans] (-2cm,7.36cm) -- (-0cm,7.27cm);

	\draw[level] (0cm,7.27cm) -- (5cm,7.27cm) node[midway,below] {$ \alpha + \alpha + \alpha  $} node[above left] {$7.27$};
\end{tikzpicture}
}
	\caption{ Population by p+$^{11}$B of the
          $ \SI{16.62}{\mega\electronvolt} $ state in
          $ {}^{12}\mathrm{C} $ and the subsequent decay
          via $ {}^{8}\mathrm{Be} $. Influence of the ground
          state in $ {}^{8}\mathrm{Be} $ is forbidden by conservation
          of angular momentum and parity. All energies are measured in MeV
          from the ground state of $ {}^{12}\mathrm{C} $.  }
	\label{fig:levelscheme}
\end{figure}
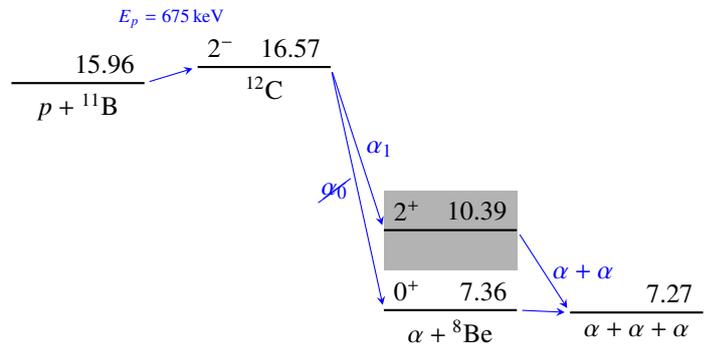

With unnatural parity s-wave partial waves cannot contribute to the
3$\alpha$-decay of the 16.62\,MeV resonance. The initial work at the
Cavendish Laboratory in the 1930s established that the 3$\alpha$-decay
is strongly influenced by the broad 2$^+$ state in $^8$Be as
illustrated in Fig.~\ref{fig:levelscheme}. Within this picture the
first $\alpha$ particle can have orbital angular momentum with either l=1
or l=3 relative to $^8$Be.
      
Until the end of the 1960s, the problem was only visited
intermittently. Then, in the period 1967-1972 at least five different
groups published detailed work on the problem. Some of this work was
spurred by MacDonald, who in 1965 pointed to the importance of
symmetry effects related to the boson character of the three $\alpha$
particles in the final state, which should lead to characteristic
patterns in the Dalitz plot of the decay~\cite{Macdonald:1965kx}.

In 1967 the first high quality data came from Kamke and
Krug~\cite{Kamke:1967qd} with a detailed analysis and interpretation
given by Sch{\"a}fer in 1970~\cite{Schafer:1970vn}, who confirmed the
presence of symmetry effects and concluded that a strong l=3 component
is needed to understand the data. In 1969 Quebert and Marquez in an
independent study also identified the symmetry effect and pointed to a
dominant l=3 contribution~\cite{Quebert:1969nr}. Next, Giorni {\it et
  al.}  published data in 1968~\cite{Giorni:1968fk} and interpretation
in 1970~\cite{Giorni:1970wd} where the data were fitted with an
adjustable coherent sum of l=1 and l=3 contributions leading to 83\%
branching ratio for l=1 (no uncertainty given). In 1970 Cockburn {\it
  et al.}  also provided high quality data and a remarkably
illuminating analysis and discussion of the
data~\cite{Cockburn:1970zr}. Cockburn {\it et al.} were able to make a
fit to the full Dalitz plot including a coherent sum of l=1 and l=3
contributions, which resulted in a ratio of reduced widths of
${\gamma_3}^2/{\gamma_1}^2$=0.60-0.65. Since the l=3 wave is
damped by the angular momentum barrier, this could be in
rough agreement with the ratio deduced by
Giorni~\cite{Giorni:1970wd}. By analysing regions in the Dalitz plot
with the largest discrepancies between theory and data, Cockburn {\it
  et al.}  could identify the effect of rescattering between primary
(emitted from $^{12}$C) and secondary (emitted from $^8$Be) $\alpha$
particles, and also examined if the properties of the $^8$Be 2$^+$
resonance were affected by the presence of the third $\alpha$
particle. The last result from this period was published by Treado
{\it et al.}~\cite{Treado:1972fk}. In this work a ratio of 10(3)
between the branching ratios to the l=3 and l=1 components in favour
of l=1 was deduced from an analysis of angular correlations between
primary and secondary $\alpha$ particles. This indicates a stronger
dominance of l=1 over l=3 relative to the findings of Cockburn {\it et
  al.} and Giorni {\it et al.}

Somewhat later two scans of the reaction $^{11}$B(p,3$\alpha$)
were published with the aim of deducing the astrophysical reaction 
rate~\cite{Davidson:1979fk, Becker:1987fk}. In these studies the treatment 
of the 3$\alpha$ breakup of $^{12}$C states populated in the reaction was 
somewhat crude, e.g. neglecting symmetrisation effects, and including only
the l=1 contribution.

In 2011 Stave {\it et al.}~\cite{Stave:2011fk} published a new measurement of 
the 3$\alpha$-decay of the 16.62\,MeV state. They refer to \cite{Becker:1987fk} 
for providing the modern view of the decay of the 2$^-$ state
emphasizing the choice of l=1. Stave {\it et al.} then use a comparison of their 
new data to a model neglecting symmetry effects to conclude that the primary
$\alpha$ particle is emitted with l=3 and the conclusions from \cite{Becker:1987fk} 
on the breakup mechanism of the 2$^-$ state therefore is erroneous.

All experiments referred to in this section used similar detector
setups consisting of one or two small detectors with one at a fixed
positions and the other free to rotate around the target position. In
contrast, the present study uses a compact array of DSSDs allowing for
a complete-kinematics measurement, as detailed later.  

\section{3$\alpha$ breakup model}
\label{model}
We use the R-matrix based description and parametrisation of the
3$\alpha$ decay from \cite{Balamuth:1974ec}. This parametrisation has
also been used to describe the decay of other states in $^{12}$C: e.g.
0$^+$ (7.65 MeV) \cite{Refsgaard:2018aa}, 1$^+$ (12.71 MeV)
\cite{Fynbo:2003ys}, and 2$^+$ (16.11 MeV) \cite{Laursen:2016ab}.
This description takes into account the symmetry stemming from the
Boson character of the $\alpha$-particles as pointed out by
MacDonald~\cite{Macdonald:1965kx}. The description is also very
similar to those used by \cite{Quebert:1969nr, Schafer:1970vn,
  Giorni:1970wd, Cockburn:1970zr}.

In this description the decay amplitude is given by
\begin{align}
\label{eq:single_amp}
f_c^{m_a}&(123) = \sum_{m_b} \langle J_b l_1 m_b (m_a-m_b) \vert J_a m_a \rangle \nonumber \\ &\times \bigl[i^{l_1} Y_{l_1}^{m_a-m_b}(\Omega_1)\bigr] \bigl[i^{l_2} Y_{l_2}^{m_b}(\Omega_{23})\bigr]  \nonumber \\ &\times  \gamma_c \bigl(2P_{l_1} / \rho_1\bigr)^{\frac{1}{2}}\exp\bigl[i(\omega_{l_1} - \phi_{l_1})\bigr] F_c(E_{23}),
\end{align}
where $F_c(E_{23})$ is a factor describing the resonant strength of
the intermediate system. In R-matrix theory the single-level
approximation is
\begin{align}
\label{eq:single_level}
F_c(E_{23}) = \frac{\gamma_{\lambda_b l_2}\bigl(2P_{l_2} / \rho_{23}\bigr)^{\frac{1}{2}}\exp\bigl[i(\omega_{l_2} - \phi_{l_2})\bigr]}{E_{\lambda_b} - E_{23} - \bigl[S_{l_2} - B{l_2} + iP_{l_2}\bigr] \gamma_{\lambda_b l_2}^2} .
\end{align}
                 
To obtain the total decay weight the expression is symmetrised by
permutation of the indices of the $\alpha$ particles:
\begin{align}
\label{eq:total_weight}
W = \sum_{m_a} \Bigl\lvert \sum_c \Bigl\lbrace f_c^{m_a}(123) + f_c^{m_a}(231) + f_c^{m_a}(312) \Bigr\rbrace \Bigr\rvert^2 .
\end{align}
The symbols appearing in eqs. \eqref{eq:single_amp}--\eqref{eq:total_weight} are defined in \ref{tab:notation}. 
\begin{table}[htbp]
\centering
\caption{Explanation of the parameters appearing in eqs. \eqref{eq:single_amp}--\eqref{eq:total_weight}.}
\medskip
\small
\label{tab:notation}
\begin{tabular}{r p{6.5cm}}
\hline 
$J_a, m_a$ & Angular momentum quantum numbers of $^{12}$C.\\
$J_b, m_b$ & Same for $^8$Be. \\
$l_1, l_2$ & Orbital angular momentum in $^{12}$C$\rightarrow$$^8$Be$+
\alpha$ and $^8$Be$\rightarrow$2$\alpha$ \\
$\lambda_b$ & The level populated in $^8$Be. Implicitly specifies
              $J_b$ and $l_2$.\\
$c$ & Decay channel specifying $\lbrace l_1, \lambda_b \rbrace$.\\
$\gamma_c$ & Reduced width amplitude for decay of $^{12}$C through channel $c$. \\
$\gamma_{\lambda_b l_2}$ & Same for decay of $^8$Be. \\
$\Omega_1$ & Direction of the first emitted $\alpha$ in the rest frame of $^{12}$C. \\
$\Omega_{23}$ & Direction of the second emitted $\alpha$ in the rest frame of $^8$Be. \\
$E_{23}$ & Relative energy between $\alpha_2$ and $\alpha_3$ \\
$\rho_1$ & $=k_1 a_1$, where $k_1$ is the wave number and $a_1$ is the channel radius for the primary breakup channel. \\
$\rho_{23}$ & Same for the secondary breakup channel.\\
$P_{l_1}, P_{l_2}$ & Penetrability for the primary and secondary breakup channels. \\
$\omega_{l_1}, \omega_{l_2}$ & Coulomb phase shifts. \\
$\phi_{l_1}, \phi_{l_2}$ & Hard-sphere phase shifts. \\
$E_{\lambda_b}$ & Level energy of $\lambda_b$ in the intermediate system. \\
$S_{l_2}, B_{l_2}$ & Shift function and boundary condition for the secondary breakup channel.    \\
\hline
\end{tabular}
\end{table}

Contributions from $l_1=1$ and $l_1=3$ can be added coherently. We will
also allow an additional adjustable phase $\delta$
between the $l_1=1$ and $l_1=3$ amplitudes. 
\begin{equation}
\left\lvert f \right\rvert^{2} = \sum_{ m_{a}} \left\lvert \sqrt{k} f_{\ell_1=1} + \sqrt{1 - k} \cdot e^{i \delta} f_{\ell_1=3} \right\rvert^{2}
\label{eq:model-interference}
\end{equation}

The question that will be adressed in the following is then: is the
3$\alpha$ breakup of the 16.62 MeV resonance well described by this
model, and if so, is the breakup dominated by $l_1=1$, $l_1=3$, or a
mixture of the two? We anticipate that the model cannot provide a
perfect description of the data since it ignores any interaction
between the $\alpha$-particle emitted in the process
$^{12}$C$\rightarrow$$^8$Be$+
\alpha$ and those from $^8$Be$\rightarrow$2$\alpha$
~\cite{Cockburn:1970zr,Fynbo:2003ys}.  

\section{Experiment}
The experiment was conducted at the $ \SI{5}{\mega\electronvolt} $ Van
de Graaff accelerator at Aarhus University. Other results from the
same experimental campaign have been published elsewhere
\cite{Munch:2018aa,Kirsebom:2020aa}.

The 16.62 MeV resonance was populated via $ {}^{11}\mathrm{B} $(p$
{}^{12}\mathrm{C}$), as depicted in Fig.~\ref{fig:levelscheme}. The
accelerator provided a $ \SI{1361(2)}{\kilo\electronvolt} $ beam of
$ H_{2}^{+}$, which was guided through a sequence of slits and
steerers into the reaction chamber where it impacted a
$ \sim \SI{120}{\nano\meter} $ $ {}^{11}\mathrm{B} $ target on a
carbon backing produced in house.  The measurement time was $ 43.1 $
hours with an integrated beam current of
$ \SI{55(3)}{\micro\coulomb} $.

The detection setup consisted of four double-sided silicon-strip
detectors (DSSSD's).  Two of the detectors were square
($ \SI{5}{\centi\m} \times \SI{5}{\centi\m} $ with 16 rectangular
strips on each side) and two were annular (inner radius
$ \SI{11}{\milli\m} $ and outer radius is $ \SI{35}{\milli\m} $ with
32 radial strips (spokes) on the front side and 24 circular strips
(rings) on the back), all from \textit{Micron Semiconductor Ltd.}


The signals from the detectors were analysed in a conventional scheme of
preamplifiers, amplifiers, ADCs and TDCs.

\section{Data analysis and discussion}

In the off-line analysis events of interest were selected by imposing energy and
momentum conservation of three detected particles,
and that the individual energies were larger than 250\,keV. Finally,
events corresponding to population of the ground state in $^8$Be,
which cannot originate from the 2$^-$ state, were identified from the
relative energy of pairs of $\alpha$-particles, and removed from
further analysis. Of the order of 6 $\times$ 10$^5$ 3$\alpha$-events
were collected in the experiment.

\begin{figure}[]
\centering
\includegraphics[width=0.85\columnwidth]{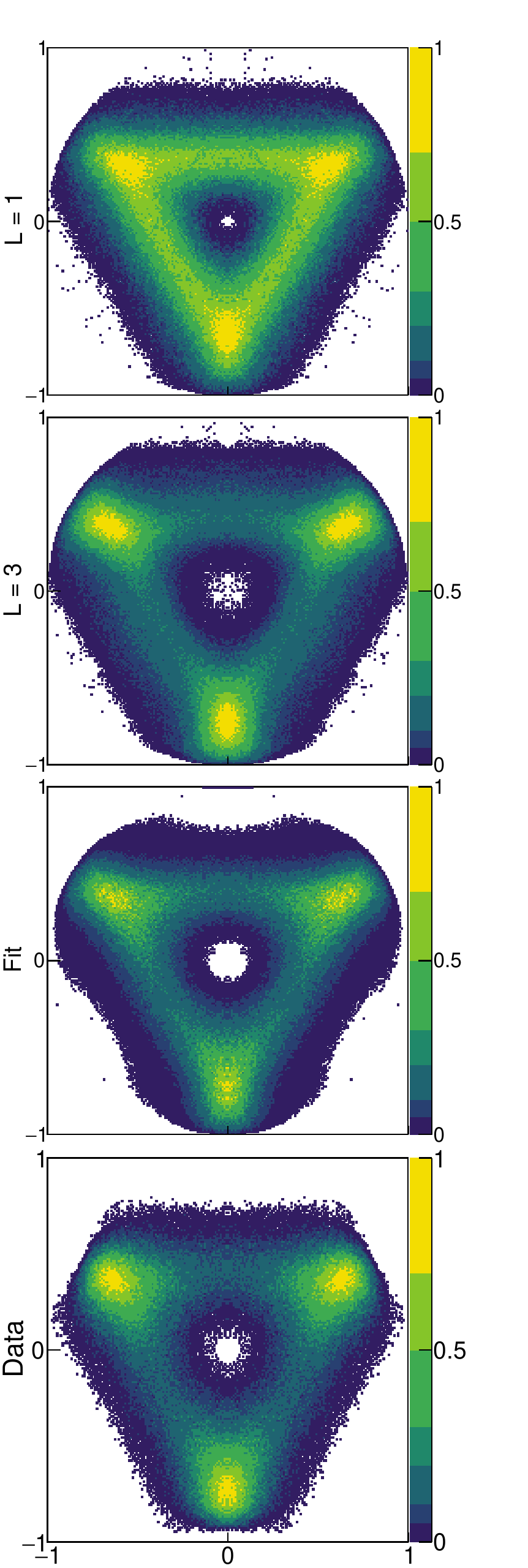}
\caption{Dalitz plots of 3$\alpha$ coincidence data and comparison to
  models with $l_1$=1, $l_1$=3 and the mixture of the two. The Dalitz
  plot coordinates are $x = \sqrt{3} (E_2 - E_3) / \sum_{i} E_{i}$ and
  $y = (2 E_1 - E_2 - E_3) / \sum_{i} E_{i}$. }
\label{fig:data_dalitz}
\end{figure}

In the lower panel in fig.\ref{fig:data_dalitz} we display the 3$\alpha$
events selected in this way in a Dalitz plot with axes
$x = \sqrt{3} (E_2 - E_3) / \sum_{i} E_{i}$ and
$y = (2 E_1 - E_2 - E_3) / \sum_{i} E_{i}$, where $E_{i}$ is the
energy of the $ i $'th alpha particle in the center-of-mass frame.

A qualitative understanding of the measured distribution of data in
the Dalitz plot comes from the general description given
in~\cite{Fynbo:2009fk}: The events are confined to the inside of a
circle due to momentum conservation. As first pointed out
by~\cite{Macdonald:1965kx}, the centre and the rim of the circle are
excluded due to symmetry requirements characteristic for a 2$^-$
state. The influence of resonances between two of the three
$\alpha$-particles is seen as lines corresponding to fixed values of
the relative energy of two $\alpha$-particles. The width of these lines
are determined by the width of the $\alpha \alpha$ resonance. Due to
the presence of three identical particles this leads to the appearance
of the triangular region of enhanced intensity in
fig.~\ref{fig:data_dalitz}. The intensity distribution along the lines
of this triangle is determined by angular correlations. Interference
effects are strongest where the lines overlap. Note, interference
effects were ignored in several studies referred to in
section~\ref{review}, e.g. \cite{Davidson:1979fk, Becker:1987fk,
  Stave:2011fk}. 

To facilitate a quantitative understanding of the data and to
elucidate the veracity of the different conclusions on the
3$\alpha$-decay of the 2$^-$ resonance outlined in
section~\ref{review} we explore if the data is well described by the
model from section~\ref{model}, and if so, if the breakup is
dominated by $l_1=1$, $l_1=3$, or a mixture of the two. The upper
three panels in fig.\ref{fig:data_dalitz} show the model prediction
for $l_1=1$, $l_1=3$, and our best fit to a mixture of the two,
respectively. In table~\ref{table:fits} we give the result of fits of
the models to the data. For completeness we also give the result of a
fit to Eq.~\ref{eq:model-interference} with $\delta$=0. Obviously, the
best model is obtained by allowing an admixture of $l_1=1$ and $l_1=3$
- this is true whether or not the extra phase $\delta$ is
included.

The fitted values of the admixture are in broad agreement with the
ratios found by~\cite{Giorni:1970wd,Cockburn:1970zr}. The two models
with only $l_1$=1 or $l_1$=3 give similar quality fits with a slight
preference for the former. This preference is somewhat surprising
based on visual inspection of the Dalitz plots in
fig.~\ref{fig:data_dalitz}.

\begin{figure}[]
\centering
\includegraphics[width=0.92\columnwidth]{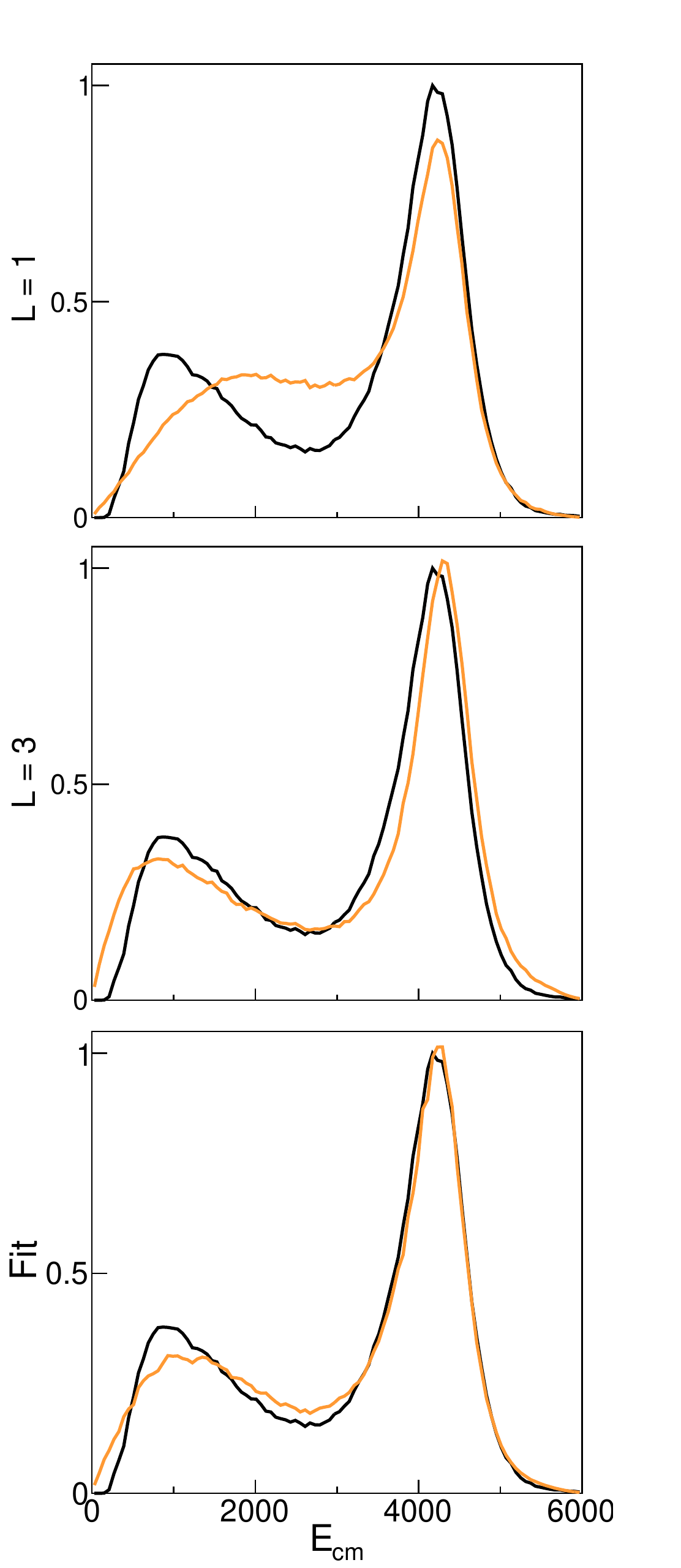}
\caption{Single $\alpha$-spectra from the same data shown in fig.~\ref{fig:data_dalitz} and comparison to models with $l_1$=1,
  $l_1$=3 and a mixture of the two. Data is black, and model is
  orange.}
\label{fig:inclusive}
\end{figure}

To understand the difference between $l_1$=1 and $l_1$=3 we show in
fig.~\ref{fig:inclusive} the projected $\alpha$-spectra from the
Dalitz-plots in fig.~\ref{fig:data_dalitz}. 
There are primarily two places where $l_1$ influences the decay
amplitude given in Eq.~\ref{eq:single_amp}: In the penetrability
$P_{l_1}$ and in the angular part through Clebsh-Gordon coefficients
and spherical harmonic functions. The effect of the penetrability can
be seen on the high energy side of the energy distribution where the
larger barrier from $l_1$=3 appears to push the main peak a little up
in energy. The difference in the low-energy part of the spectra is due
to the difference in angular correlations for $l_1$=1,3; the latter
effect is discussed in more detail in
\cite{Fynbo:2003ys,Laursen:2016ab}. The angular part also accounts for
the difference in the size of the hole in the centre of the Dalitz-plot,
which corresponds to events with equal energies of the three
$\alpha$-particles. This corresponds to individual energies around
3~MeV. 

Allowing for an admixture of the two angular momenta gives the best
fit because all of these features can be better reproduced. Allowing
the $^8$Be 2$^+$ resonance energy to move by more than 100keV does not
change the conclusion that $l_1$=1 dominates when the two values are
both allowed to contribute.

Note, the energy spectra displayed in fig.~\ref{fig:inclusive} are
very similar to those in~\cite{Quebert:1969nr} and to some extent also
those in~\cite{Stave:2011fk}. To further elucidate the conclusions
from previous studies we have also compared our data to a model
neglecting interference effects, which is obtained by using only one
of the three permutations in Eq.~\ref{eq:total_weight}. This model is
equivalent to those used by \cite{Davidson:1979fk, Becker:1987fk,
  Stave:2011fk}. With this model, $l_1$=3 gives the best reproduction
of our $\alpha$-spectrum in agreement with the conclusion of
\cite{Stave:2011fk}. It therefore appears that the conclusion of the
dominance of $l_1$=3 is based on using too simple a model that only
allows for one value of $l_1$ and neglects interference, and basing
the conclusion primarily on inclusive data.

\begin{table}[htpb]
\centering
\begin{tabular}{rccrc}
\toprule
	 & $ \ell_1 = 1 $ [\%] & $ \ell_1 = 3 $ [\%]& $ \chi^{2} $ & $\delta$ \\ \midrule
	 & $ 100$ & $ 0 $ &  22705 & n/a \\
	 & $ 0$  & $ 100 $ & 22808 & n/a \\
	 & 85(1) & 15(1) & 15360 & $ 0 $ \\
	 & 76(5) & 25(5) & 12297 & $ 67(1)\% \cdot 2 \pi  $\\
\end{tabular}
\caption{ Fit results and $ \chi^{2} $ values for the tested models.
  Uncertainties are statistical only. The first two columns give the
  percentage of $\ell_1=1$ and $\ell_1=3$.  The number of degrees of
  freedom for the $ \chi^{2} $-distribution is $ \sim 5350 $.  }
\label{table:fits}
\end{table}


As seen from the values of $\chi^2$ in table~\ref{table:fits}, even our
best model deviates significantly from the data. Those differences are
presumable caused by final state interactions as discussed in some
detail~\cite{Norbeck:1968aa, Cockburn:1970zr,Thompson:1972aa,
  Fynbo:2003ys}. This aspect will not pursued further here.

\section{Conclusion}
In summary, we have collected exclusive data from the 3$\alpha$-decay
of the 16.62\,MeV (2$^-$, T=1) resonance in $^{12}$C and compared our
measurements with R-matrix based sequential models. We found that the
model with admixture of $l_1$=1 and 3 describes the data quite
well. We also demonstrate that the conflicting conclusions from
previous measurements spanning almost a century may be partly
explained by neglecting the possible admixture or two angular momenta,
neglecting interference effects or basing conclusions only on
inclusive data.

It remains to be examined what the consequences are of our results on
e.g. the rate of the $^{11}$B(p,3$\alpha$) and therefore on scenarios
where this rate plays a role including astrophysics and possibly fusion
reactors.


\bibliography{bib_desk3}

\end{document}